# FEL OSCILLATORS WITH TAPERED UNDULATORS: INCLUSION OF HARMONIC GENERATION AND PULSE PROPAGATION

G. DATTOLI, P. L. OTTAVIANI and S. PAGNUTTI

ENEA

V. ASGEKAR

Department of Physics, University of Pune, Pune 411007, INDIA

ABSTRACT

We review the theory of FEL oscillators operating with tapered undulators. We consider the case of a uniform tapering and introduce a parameter which characterizes the effect of the tapering on the gain and on the saturation intensity. We analyze the effect of the tapering on the FEL dynamics by including the pulse propagation effects too. We analyze the importance of tapering as a tool to model the optical pulse shapes and to control the higher harmonic intensities.

## 1 INTRODUCTION

FEL oscillators operating with tapered undulators have been discussed in the past, but the relevant theory and phenomenology require some clarification,s because there are some not fully understood aspects, which deserves further consideration. Furthermore the problems associated with the pulse propagation effects and non linear harmonic generation have not been discussed in depth and, as we will report here, they give rise to new and interesting dynamical features which are worth to be studied carefully.

Originally the concept of undulator tapering was introduced for FEL amplifiers [1] and its straightforward extension to the oscillator regime has been the source of some surprises, regarding the relevant consequences on the oscillator efficiency.

In the case of the amplifier the tapering is usually designed in such a way that the undulator field decreases (or the period as well) in the forward direction, in order to compensate the effect of the energy losses of the e-beam and ensure an efficient trapping of the electrons in a stable bucket. Reverse tapering is simply the contrary, namely increase of the field, which causes the reduction of the FEL amplifier efficiency. On the other side a mild reverse tapering may be a tool to enhance the efficiency in the case of FEL oscillators [2-4]. The problem of FEL oscillator with tapered undulators has been the topic of experimental and theoretical investigations, but some interesting aspects have not been explored yet, such as the effect of tapering on the non linear harmonic generation. Here we will fill some of these gaps and consider quite a general treatment by including small and strong signal regimes. The analysis will be however limited to linear tapering and we will not develop strategies for an optimal tapering, which is a not well defined concept within the framework of the present investigation.

This paper combines analytical and numerical results and it is also devoted to the derivation of practical formulae, which can include the effect of the tapering to the gain and to the saturation intensity. The paper consists of two parts.

In the first we will establish practical formulae concerning the FEL operation with a linear tapering (namely with undulators exhibiting an on axis field amplitude depending linearly on the longitudinal coordinate). In particular we will derive the gain dependence vs. the tapering depth, the gain saturation formula, the saturation intensity and the efficiency factor vs. tapering depth.

In the second part we will discuss the pulse propagation effects and the interplay between slippage, short pulses and tapering. We will therefore investigate how the tapering combines with slippage and lethargy effects, to give rise to a new and interesting phenomenology.

## 2 Small signal gain including tapering

In this section we will discuss different level of approximation concerning the tapered FEL phenomenology. We will indeed consider the small signal low gain effects and we will essentially recover the results of ref. [2], we will then include the high gain corrections and the consequences of the pulse propagation effects.

Following ref. (2) we write

$$G_T(\nu, \mu_T) = -2\pi g_0 \operatorname{Im}(g(\nu, \mu_T)),$$
$$g(\nu, \mu_T) = \int_0^1 d\tau \int_0^\tau d\xi\, \xi\, e^{-i\nu\xi + i\frac{\pi \mu_T}{2}\xi(2\tau - \xi)} \qquad (1)$$

where

$$\mu_T = 2N \frac{\Delta B}{B} \frac{K_0^2}{1+\frac{K_0^2}{2}},$$

$$\Delta B = B(L_u) - B(0)$$

(2)

$N$ is the number of undulator periods, $L_u = N\lambda_u$ is the undulator length, $\mu_T$ is the tapering parameter, $\frac{\Delta B}{B}$ is the field variation along the undulator, having a uniform tapering and $K_0$ is the field strength at the undulator entrance[1].

The physical meaning of the $\mu_T$ parameter should be understood as follows. The phase matching condition, after any undulator period advance, requires that the electron longitudinal velocity, the undulator period and the FEL wavelength be related by $(1-\beta_z)\lambda_u = n\lambda$ to ensure constructive interference. In the case of tapered devices the longitudinal component of the velocity depends on the position inside the undulator, namely $\beta_z \cong 1 - \frac{1}{2\gamma^2}(1+\frac{K^2(z)}{2})$, therefore the matching condition cannot be satisfied. This fact determines a phase mismatch, which can be associated with the frequency shift, induced by the tapering. By assuming linear tapering we can set $K(z) = K_0(1+\alpha z)$ where $\alpha = \frac{\Delta B}{B} \ll 1$, which yields a detuning shift just given by $\delta \nu \cong 2\pi \mu_T$. This factor plays a role not dissimilar from the inhomogeneous broadening effects and indeed it determines a shift of the maximum gain position and a gain reduction.

As already remarked the condition of positive $\alpha$ is referred to in literature as "inverse tapering" to denote the fact that the strength of the field increases with increasing z. The gain vs. the frequency detuning $\nu$ is shown in Figs. 1-3 for different values of the tapering parameter.

---

[1] Eq. (2) implies an undulator field only undergoing a variation along the longitudinal direction; more in general the undulator period can be tapered too, in this case $\frac{\Delta B}{B}$ in eq. (2) should be replaced by $\frac{\Delta K}{K}$.

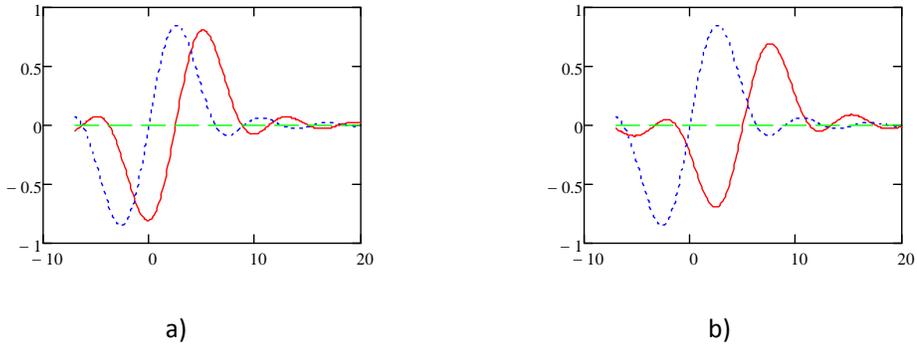

**Fig. 1** Gain curves vs. frequency detuning. a) continuous line $\mu_T = 5/\pi$, dot line $\mu_T = 0$, b) same as a) $\mu_T = 10/\pi$

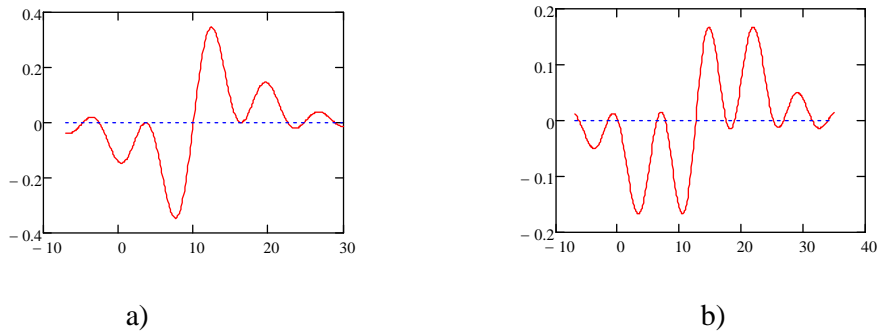

**Fig. 2** Gain curves vs. frequency detuning: a) $\mu_T = 20/\pi$, b) $\mu_T = 25.2/\pi$

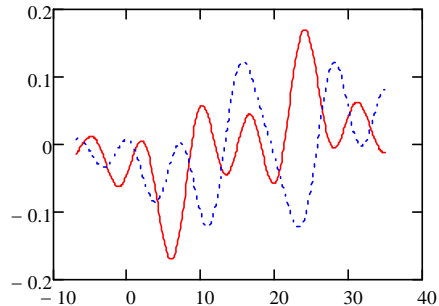

**Fig. 3** Gain curves vs. frequency detuning: continuous $\mu_T = 30/\pi$, dot $\mu_T = 39/\pi$

It is evident that when $\mu_T$ is small, the gain curve resembles, apart from a shift in the detuning parameter, the ordinary low gain curve. When the tapering parameter increases, the maximum gain shifts, smoothly, towards larger detuning. Further increase corresponds however to an abrupt variation of the maximum gain position. The threshold for the occurrence of this effect is $|\pi \mu_T| \cong 25.2$.

The value of the detuning, corresponding to the maximum gain and containing the corrections due to the tapering, is provided by the following simple relation

$$v^*(\mu_T) \cong v^*(0)[1 + 0.1844\pi\mu_T], \qquad |\pi\mu_T| < 25.2, \quad g_0 \leq 0.3 \tag{3}$$

shown in Fig. 4, where we have reported $v^*$ vs. $\pi\mu_T$.

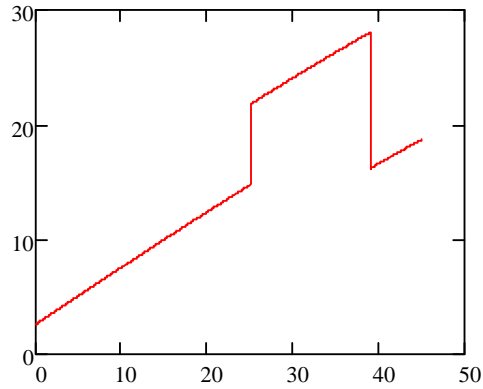

**Fig. 4** $v^*$ vs $\pi\mu_T$ ; the plot includes the region above the instability threshold

For values of the tapering parameter larger than 25.2 the curve exhibits a kind of discontinuity due to a longitudinal mode flipping, whose consequences will be commented later in the paper. The maximum gain vs. the tapering parameter is reported in Fig. 5, where it is also evident the effect of discontinuity in the detuning parameter.

The gain function for $|\pi\mu_T| \leq 25.2$ is reproduced by the equation

$$G_M(\mu_T) \cong G_M(g_0) P(\mu_T),$$

$$G_M(g_0) \cong 0.848 g_0 \tag{4a}$$

$$P(\mu_T) = 1 + 0.275\left[1 - \frac{2}{1 + e^{0.301(\pi\mu_T)}}\right] - 0.0428\pi\mu_T$$

For larger values of the tapering parameter the gain is almost constant and exhibits the following parabolic form ( $25.3 \leq \pi|\mu_T| \leq 39$ )[2]

$$G_M(\mu_T) \cong 0.168 \cdot g_0 Q(\mu_T)$$
$$Q(\mu_T) = 1 + 13.69 \cdot 10^{-3}(\pi\mu_T - 25.3) - 25 \cdot 10^{-4}(\pi\mu_T - 25.3)^2$$
(4b)

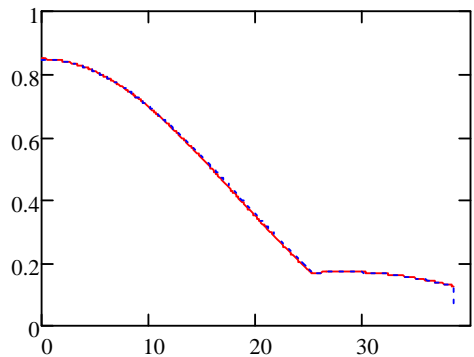

**Fig. 5** $G_M/g_0$ vs. $\pi\mu_T$; the dot line is the fitting curve

The results obtained so far confirms those of ref. [2] and show that the role of tapering is fairly more complicated than usually believed. The gain exhibits indeed a rich structure, which, as we will see in the following, is responsible for an unexpected and interesting dynamical behavior of FEL oscillators operating with non constant parameter undulators.

The considerations we have developed so far do not include high gain effects, which can be accounted for using the integral equation

$$\partial_\tau a = i\pi g_0 \int_0^\tau d\xi\, \xi\, e^{-i\nu\xi + \frac{i\pi\mu_T}{2}\xi(2\tau-\xi)} a(\tau-\xi)$$
(5a)

It is well known, from the ordinary FEL gain theory, that, when the small signal gain coefficient increases, the gain curve loses its anti-symmetric shape, an example is in Fig. 6a where we have reported the gain of a FEL operating with a large small signal gain ($g_0 = 2$) and a modest tapering parameter. This means that, for increasing $g_0$, non linear contributions in the gain coefficient play an increasingly important role. It has however been shown that these high gain contributions tend to disappear, when inhomogeneous broadening contributions (due to energy spread and/or emittance), are active [5, 6]. This effect occurs with the increase of the tapering parameter too, as shown in Fig. (6).

---

[2] We will not consider values of the tapering above the first threshold because the gain becomes too small and the effects of the tapering have no practical importance. It was however unexpected to find the occurrence of a further peak shift and for this reason we have quoted such extreme values.

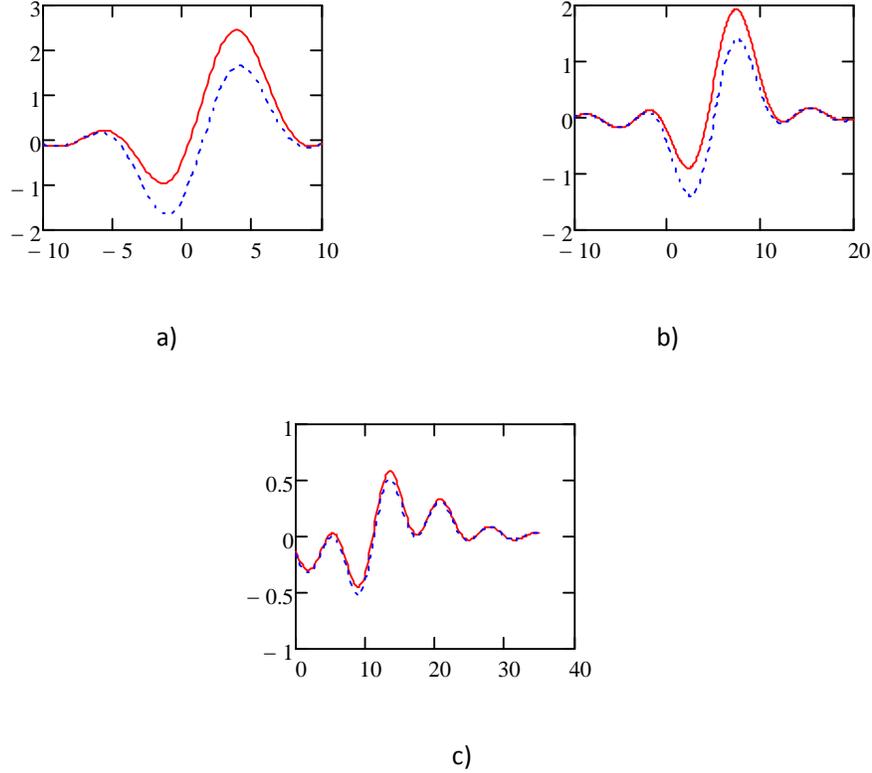

**Fig. 6** Gain vs. detuning, continuous line with high gain corrections, dot without high gain corrections;   a) $g_0 = 2, \mu_T = \dfrac{3}{\pi}$ ;  b) $g_0 = 2, \mu_T = \dfrac{10}{\pi}$ ;  c) $g_0 = 2, \mu_T = \dfrac{22.5}{\pi}$.

We have checked the validity of the previous gain curves by evaluating the gain dependence vs. the detuning by means of the simulation code Prometeo [5] (the code has been run with the parameters reported in Tab. I). The comparison, discussed below, confirms the correctness of the previous analysis.

| | |
|---|---|
| $E[MeV]$ | 155.3 |
| $\lambda_u[cm]$ | 2.8 |
| $K_0$ | 2.13 |
| $N$ | 50 |

| Rel. Energy spread | $10^{-4}$ |
| --- | --- |
| e-bunch length $\mu m$ | $10^2$ |
| Cavity length [m] | 2 |

**Tab. I** Simulation Parameters

For large values of the small signal gain coefficient the previous formulas for the value of the detuning $v^*$ corresponding to the maximum gain and for the maxim gain $G_M$ need slight corrections in $g_0$, reported below:

$$v^*(g_0, \mu_T) \cong v^*(g_0, 0)\left[1 + 0.1832(1 + 0.0631 g_0)\pi \mu_T\right],$$

$$G_M(g_0, \mu_T) \cong G_M(g_0) P(\mu_T, g_0),$$

$$G_M(g_0) \cong 0.848 g_0 + 0.19 g_0^2 \tag{5b}$$

$$P(g_0, \mu_T) = 1 + 0.275\left[1 - \frac{2}{1 + e^{0.301(\pi \mu_T)}}\right] - 0.0428(1 + 0.042 g_0)\pi \mu_T, \quad g_0 < 2$$

In Fig. 7, we have reported the gain vs. $\pi \mu_T$ for different values of $g_0$.

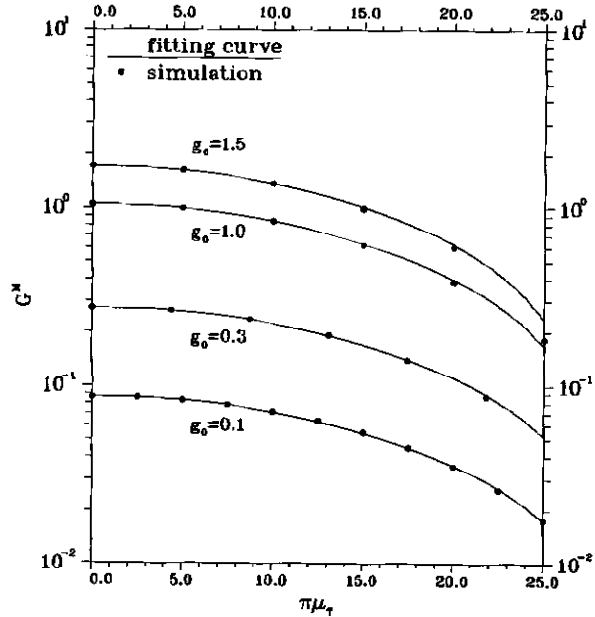

**Fig. 7** Gain vs. $\pi \mu_T$ for different values of $g_0$; the dots are the result of the simulation, the continuous curve is the fitting curve (eq. 5b)

Reversing the sign of the tapering does not create significant difference (at least in the small signal regime), as shown in Fig. 8, where gain curves with opposite tapering have the same shape, but their maxima are located at $v*(0)[1-0.1844|\pi\mu_T|]$.

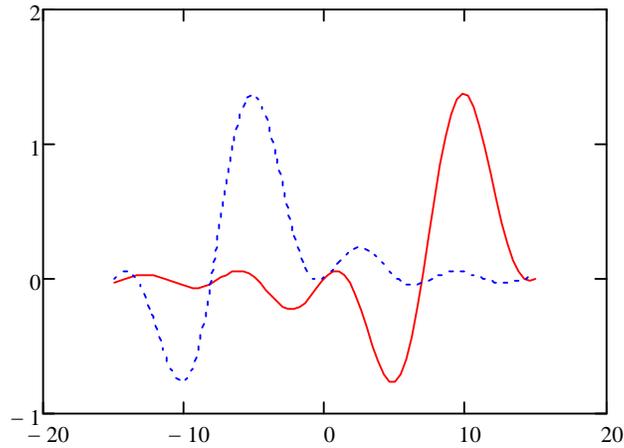

**Fig. 8** Gain vs. the detuning for $g_0 = 2$;
continuous line $\pi\mu_T = 15$, dot line $\pi\mu_T = -15$

Before concluding this section it is worth adding some comments on the agreement between the gain obtained with the numerical code and the integral equation: see e.g. Fig. 9. The disagreement between analytical and numerical solution for high $\mu_T$ is just due to the fact that the analytical procedure does not provide any corrections associated with the variations of the *K* parameter, in the small signal gain coefficient. The relative error associated with the Bessel factor term in the gain coefficient is proportional to $\left(\frac{\Delta K}{2K}\right)^2$ and becomes therefore more significant with the increasing of the tapering parameter.

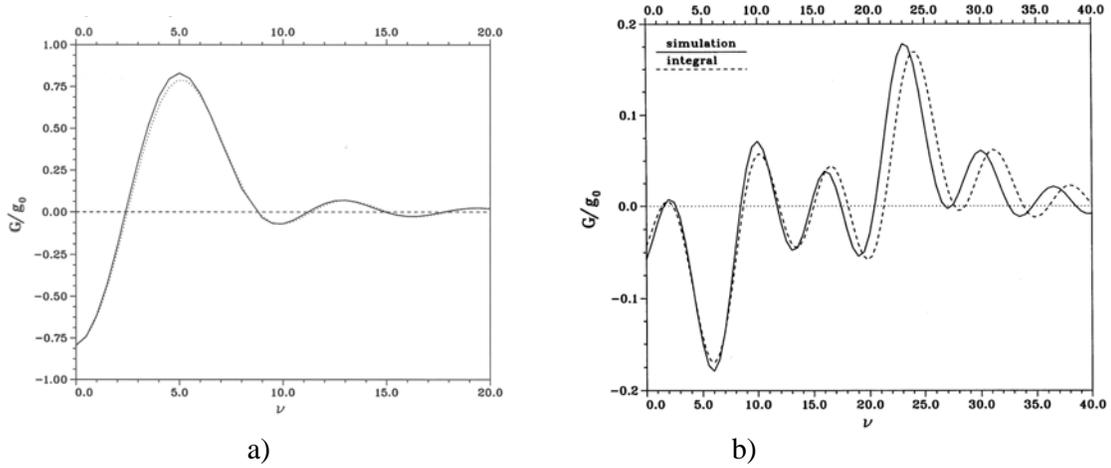

a)                                        b)

**Fig. 9** Gain curve vs. detuning; dot line: analytical formula; continous line: simulation

$$a)\ g_0 = 0.1,\ \pi\mu_T = 5,\quad b)\ g_0 = 0.1,\ \pi\mu_T = 30$$

### 3 The Saturation intensity

In the previous section we have dealt with the small signal regime, here we will introduce the saturation effects and discuss how other quantities of crucial importance for the FEL, like the saturation intensity, are affected by the tapering.

We remind that we mean, by saturation intensity, the value of the laser intensity halving the FEL small signal gain.

The gain saturation mechanism consists essentially of two parts: one is a kind of frequency shift towards the negative part of the gain and the other is an effect of inhomogeneous broadening, associated with the induced energy spread.

Let us introduce the FEL Hamiltonian which can be quite useful to understand the previously quoted points. We write, according to ref. [7], the tapered FEL Hamiltonian as

$$H = \frac{1}{2}v^2 + \pi\mu_T\zeta - |a|\sin(\zeta + \phi) \qquad (6)$$

where the tapering contributes to the FEL dynamics like a kind of accelerating electric field. The above Hamiltonian is the rigorous formulation of the already quoted effect of the tapering on the FEL dynamics.

The fact that the gain is shifted towards larger $v$ values implies that it has larger kinetic energy to overcome the trapping potential. For this reason the saturation intensity should be larger than the corresponding value of the non tapered case.

In Fig. 10 we have reported the saturation intensities vs. tapering parameter for different $g_0$; the relevant behavior is reproduced by

$$\frac{I_s(g_0,\mu_T)}{I_s(g_0,0)} = 1 + 0.017\,\pi\,\mu_T + 0.162(e^{4.8\cdot 10^{-3}(\pi\mu_T)^2} - 1), \qquad g_0 < 2 \qquad (7)$$

The saturation intensity $I_s(g_0,0)$ contains the high gain corrections too (for an analytical formula see [5]). We stress that eq. (7) is valid for reverse tapering only. In the case of direct tapering (negative values of $\pi\mu_T$ the situation is different and the associated saturation mechanisms will be discussed elsewhere).

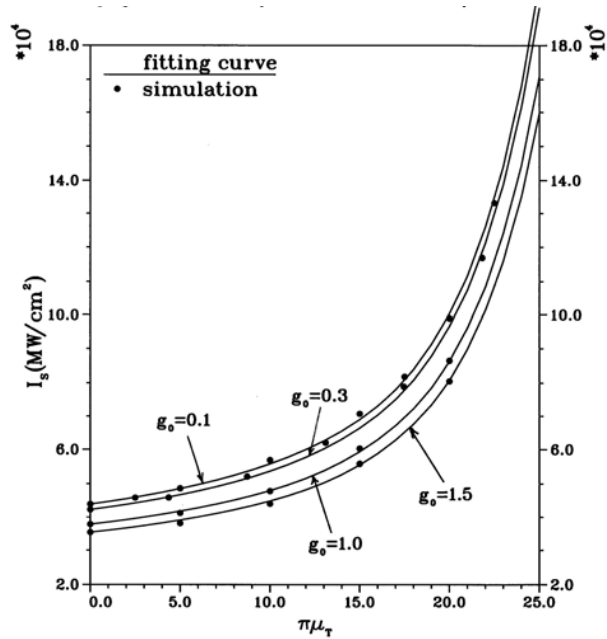

**Fig. 10** Saturation intensity vs. tapering parameter for different $g_0$; the dots are the result of the simulation, the continuous curve is the fitting curve (eq. 7)

We can combine gain formula and saturation intensity to get the round trip, $r_t$, evolution of the intracavity field $I(r_t)$. The result can be expressed in terms of the discrete logistic equation [8], reported below

$$I(r_t) = I_0 \frac{[(1-\eta)(G_M(g_0,\mu_T)+1)]^{r_t}}{1 + \frac{I_0}{I_e(g_0,\mu_T,\eta)}\left\{[(1-\eta)(G_M(g_0,\mu_T)+1)]^{r_t} - 1\right\}}, \quad (8)$$

$$I_e(g_0,\mu_T,\eta) = (\sqrt{2}+1)h(\mu_T,\eta/g_0)\left[\sqrt{\frac{1-\eta}{\eta}}G_M(g_0,\mu_T) - 1\right]I_s(g_0,\mu_T),$$

$$h(\mu_T,\eta/g_0) = 1 + a(\mu_T)e^{-b(\mu_T)\frac{\eta}{g_0}} \qquad \pi\mu_T < 20, \quad \frac{\eta}{g_0} > 0.04$$

$$a(\mu_T) = 0.124 + 0.063\, e^{0.262\pi\mu_T}$$

$$b(\mu_T) = 9.13 - 0.34\pi\mu_T + 0.068(\pi\mu_T)^2$$

where $I_0$, $I_e(g_0,\mu_T,\eta)$ denote the input seed and the equilibrium intracavity intensities respectively. The function $h(\mu_T,\eta/g_0)$ is an ad hoc introduced correction, which accounts for the high intracavity equilibrium power, occurring at low cavity losses. The comparison between fitting formula and numerical results for $I_e$ is reported in Fig. 11.

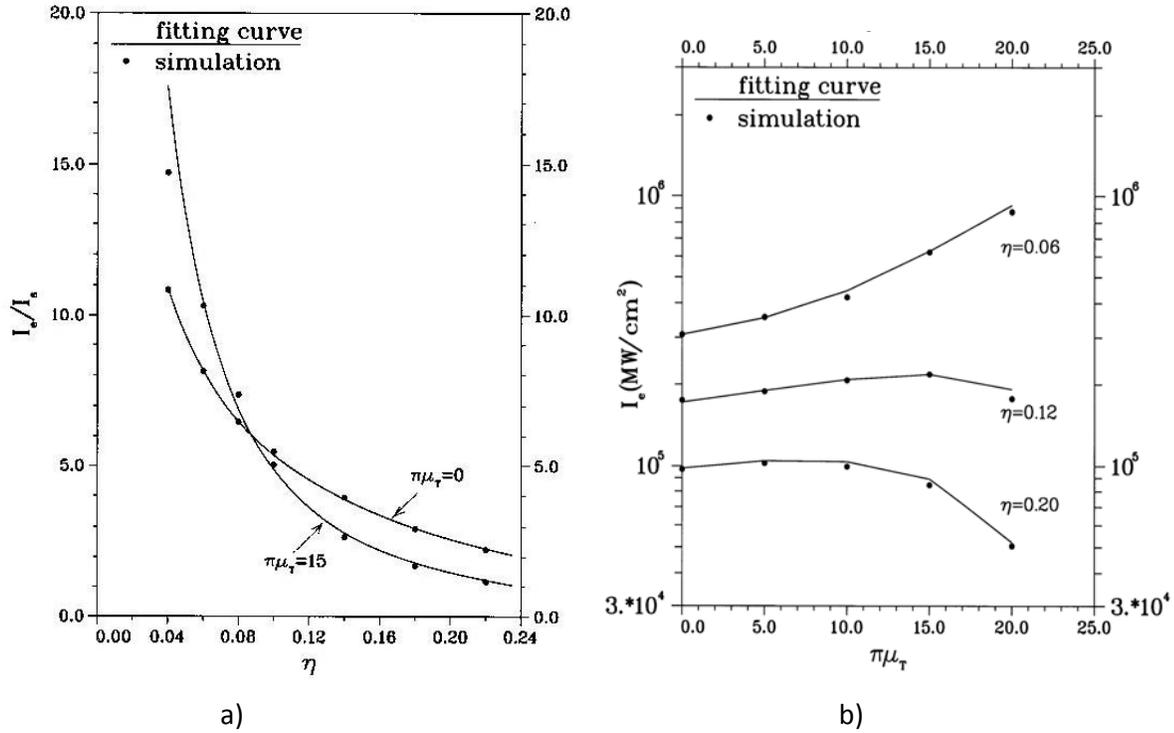

**Fig. 11** Intracavity equilibrium power; comparison between numerical results (dots) and fitting formula (continuous line) for $g_0 = 1$: a) vs. cavity losses, b) vs. tapering parameter

We can distinguish passive $\eta_P$ and active $\eta_A$ cavity losses such that

$$\eta = \eta_A + \eta_P = \eta_A(1+r), \qquad r = \frac{\eta_P}{\eta_A}$$

$$I_{out}(\mu_T, \eta_A, r) = \eta_A I_e(\mu_T, \eta)(1 + G_e) = \frac{\eta_A}{1-\eta} I_e(\mu_T, \eta)$$

(9)

where $G_e = \frac{\eta}{1-\eta}$ is the equilibrium gain.

The output power is therefore just given by the product of the active losses time the intracavity power density. In Fig. (12 a,b) we have reported the dimensionless output power vs. the active losses for different values of the tapering parameter. Fig. 12b shows the simulation results obtained with Prometeo, while Fig. 12a reports the result of a semi-analytical computation based on previous formulae.

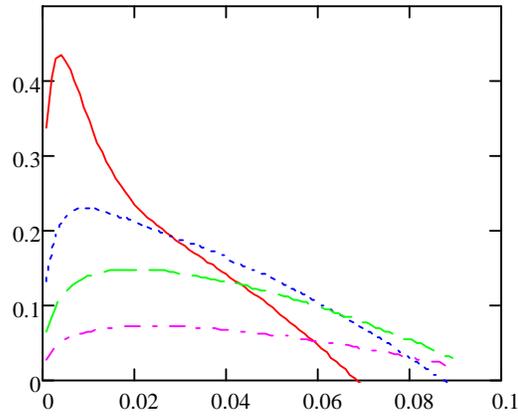

**Fig. 12a** Dimensionless output power density vs. $\eta_A$, $r=0.4$, $g_0=0.2$;
$\pi\mu_T=15$ continuous line, $\pi\mu_T=10$ dot line, $\pi\mu_T=5$ dash line, $\pi\mu_T=0$ dash-dot line

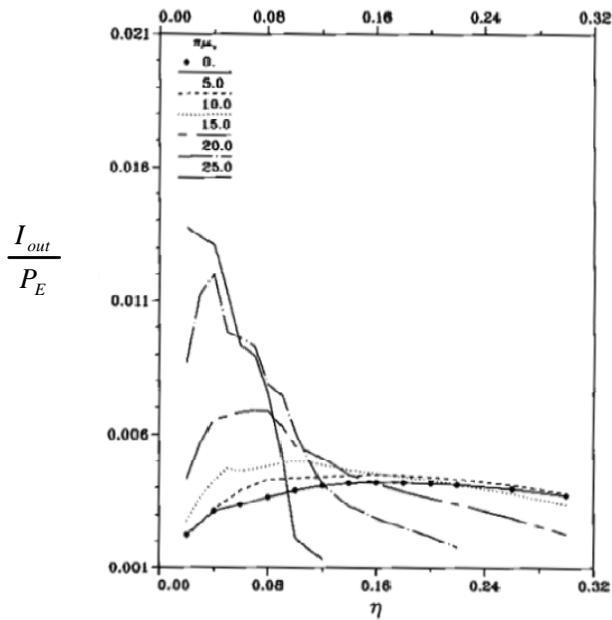

**Fig. 12b** Ratio of output power to electron power vs. losses
for different $\pi\mu_T$; $r=0.4$, $g_0=1$

The conclusion we may draw from Fig. 12 is that the optimum value of the cavity losses depends on the value of the tapering; this is not surprising since the tapering affects the peak gain. It must however be stressed that present results are fairly accurate but they are the consequence of a one dimensional analysis, the inclusion of the transverse optical mode structure could induce some changes, which do not hamper the conclusion of the paper.

Let us now discuss whether a tapered ubdulator determines an effective enhancement of the FEL efficiency, in the case of the oscillator configuration. To this aim, we remind that the efficiency is defined as the ratio between the FEL output power density and electron beam power density $P_E$, namely

$$E(\mu_T) = \frac{I_{out}(\mu_T, \eta_A, r)}{P_E}$$

$$E(0) = \frac{I_{out}(0, \eta_A, r)}{P_E} \qquad (10)$$

The efficiency enhancing factor is

$$e(\mu_T, \eta_A, r) = \frac{I_{out}(\mu_T, \eta_A, r)}{I_{out}(0, \eta_A, r)} = \frac{E(\mu_T)}{E(0)} \qquad (11).$$

An effective increase of the efficiency due to the tapering occurs therefore whenever $e(\mu_T, \eta_A, r) > 1$.

In Fig. 13 we have reported the efficiency enhancing factor vs. the tapering parameter; it is evident an effective increase[3] for values of $\pi \mu_T$ above 10.

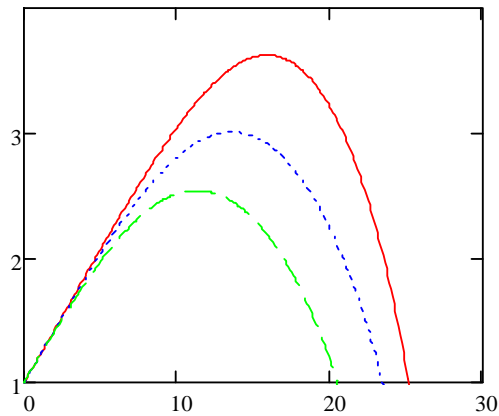

---

[3] Note that Fig. 13 shows the increase of efficiency with respect to the value with zero tapering parameter. It does therefore not imply that the optimum of the operation is at 3% cavity losses, but only that for this value of the active losses the tapering is more efficient.

**Fig. 13** Tapered FEL efficiency enhancing factor vs. $\pi \mu_T$ for different values of the active losses, $g_0 = 0.3$, $r = 0.2$; $\eta_A = 3\%$ continuous line; $\eta_A = 4\%$ dot line; $\eta_A = 6\%$ dash line

This conclusion is consistent with the experimental results of refs. [ 4].

An accurate comparison with these results requires a full three dimensional analysis, because the quoted experiments have been performed in the FIR with a tapered FEL oscillator, operating with a wave guide and an hybrid cavity with an output toroidal holed mirror.

## 4 Pulse propagation effects

This concluding section is not dedicated to the efficiency problem but to the phenomenology emerging from the inclusion of pulse propagation effects in tapered FEL oscillator [9]. It is particularly interesting because, as we will see in the following, new interesting dynamical features emerge which make the uniform tapering an additional tool to control the laser beam quality.

The equation yielding the small signal evolution of the FEL field with the inclusion of tapering and short pulses can be written as

$$\partial_\tau a = i\pi g_0(z+\Delta\cdot\tau)\int_0^\tau d\xi\, \xi\, e^{-i\nu\xi + \frac{i\pi\mu_T}{2}\xi(2\tau-\xi)} a(z+\Delta\cdot\xi, \tau-\xi)$$

$\Delta = N\lambda \equiv slippage\ length,$

$g_0(z) = g_0 f(z),$ (12)

$f(z) \equiv electron\ packet\ shape$

The slippage length is due to the different velocities between electrons and radiation, which causes a slipping of the optical bunch over the electrons during the interaction inside the undulator. The combination of tapering and slippage give rise to a new phenomenology.

Albeit we will discuss these aspects of the problem in a dedicated paper, here we will present a few interesting results emerging from the numerical treatment of the problem. In Fig. 14 we have reported the evolution vs. the round trip of the optical pulses, for the case with zero, negative and positive values of the tapering parameter and the cavity set at zero mismatch. The sequence shows that, for the same small signal gain coefficient, the effect of the tapering is an obvious reduction of the small signal gain which determines an increase of the time necessary to reach the saturation (see also Fig. 15) and, for non zero values of the tapering parameter, a kind of compensation of the slippage effect: the radiation packets tend to be more confined within the electron packet; this effect is more evident for reverse tapering.

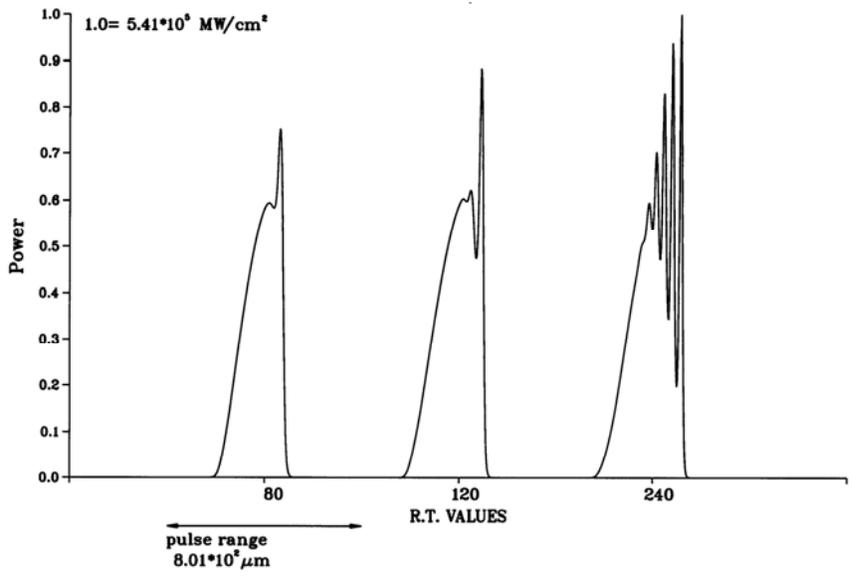

Fig. 14  a)

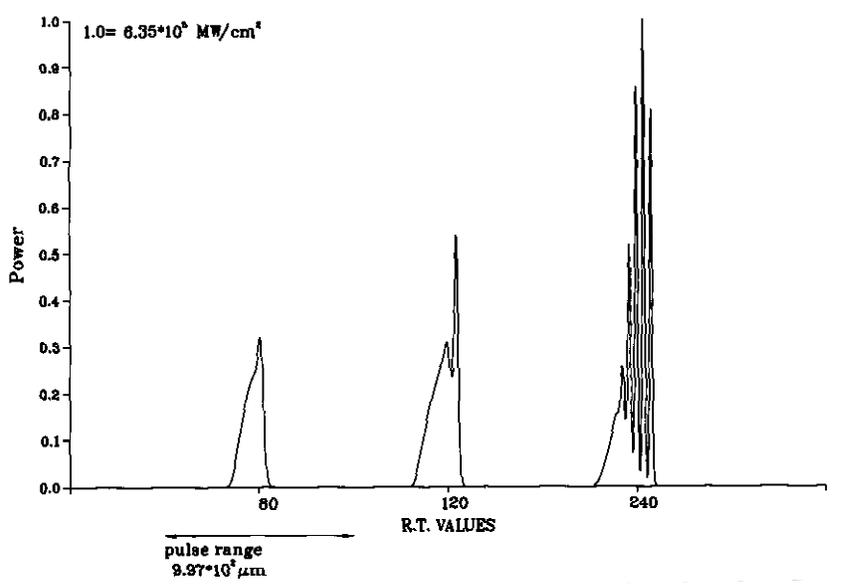

Fig. 14  b)

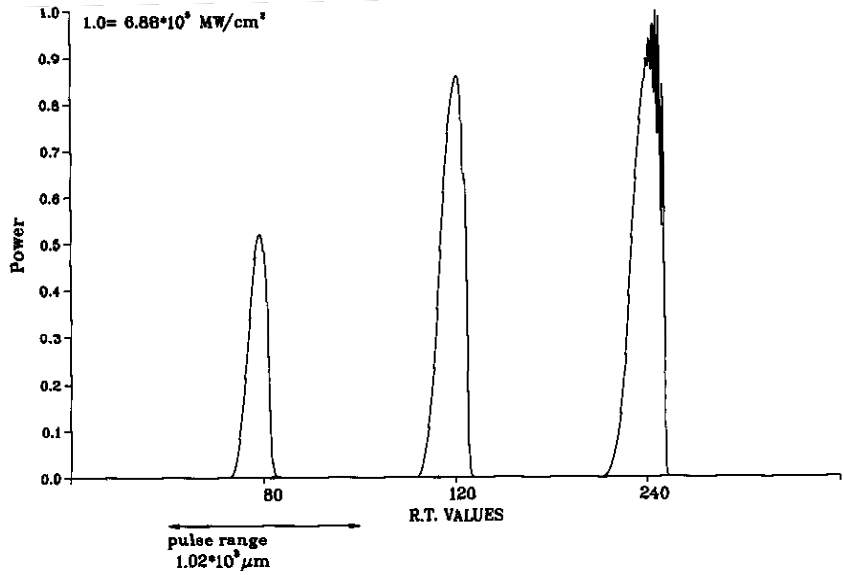

**Fig. 14** c)

**Fig. 14** Optical pulses vs. round trip for $g_0 = 1$, $\eta = 0.06$ and zero cavity mismatch;

$$a)\, \pi\, \mu_T = 0, \quad b)\, \pi\, \mu_T = -15, \quad c)\, \pi\, \mu_T = 15$$

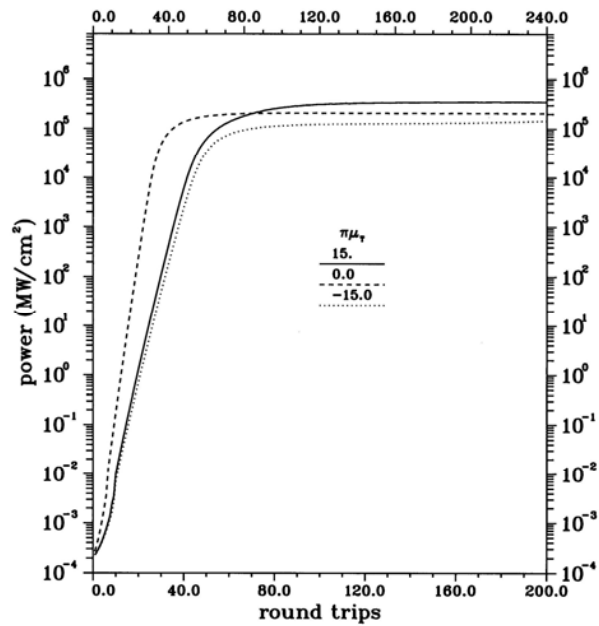

**Fig. 15** Intracavity power evolution vs. round trip number for different $\pi\, \mu_T$, same parameters as Fig. 14

In Fig. 16 we have detailed the pulse shapes of the previous snap shots in the region of deep saturation. It is interesting to note that a kind of comb-like structure occurs well beyond saturation. For the parameters we have considered in the simulation, saturation is reached after 80 round trips and we have run the code up to 240 round trips.

The physical origin of the peaks is essentially due to a strong mode locking mechanism occurring at the scale of the cooperation length. The effect is a genuine combination between mode-locking and slippage effects. The longitudinal mode locking occurs because we are considering a FEL oscillators in which the mode coupling (active mode locking) is induced by the pulsed structure of the bunch itself. The effect was predicted long ago in ref. [10] and is not dissimilar from the mechanism determining the occurrence of the spikes in high gain SASE FEL devices [11]. Further comments on this last points can be found in ref. [12].

This effect is also evident for the pulses associated with the higher order non linear harmonics, which, in the tapered case, has a less defined structure due to the fact that the optical pulse well overlapped to the electron packet is creating a strong bunching in a region spread over all the electron bunch.

The consequence of this type of structure on the coherently generated harmonics is shown in Fig. 17 where we have reported the third harmonics generated with no tapering and inverse tapering; the figures put in evidence the sensitivity of the harmonic pulse shape to the tapering, which becomes a tool to shape the emerging pulse.

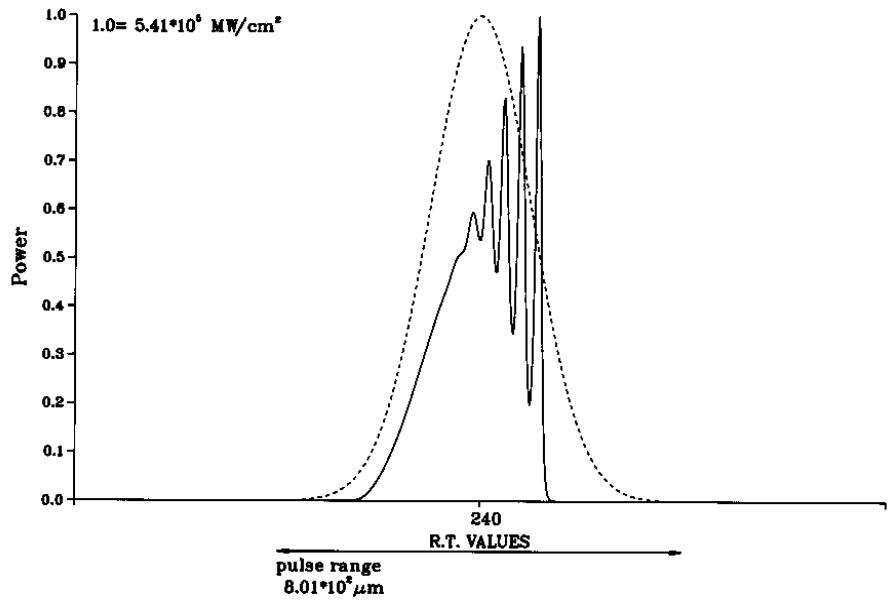

**Fig. 16 a**

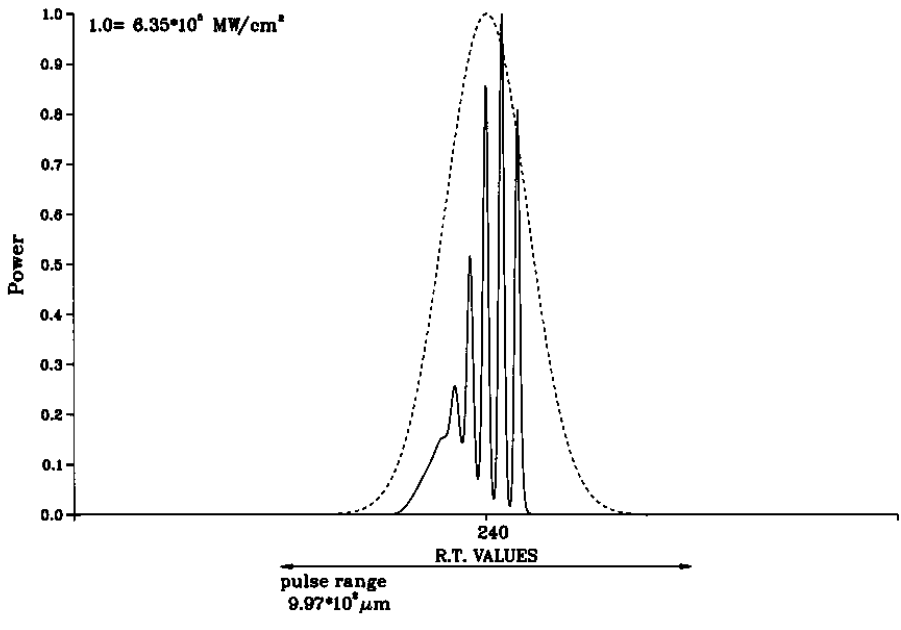

**Fig. 16 b**

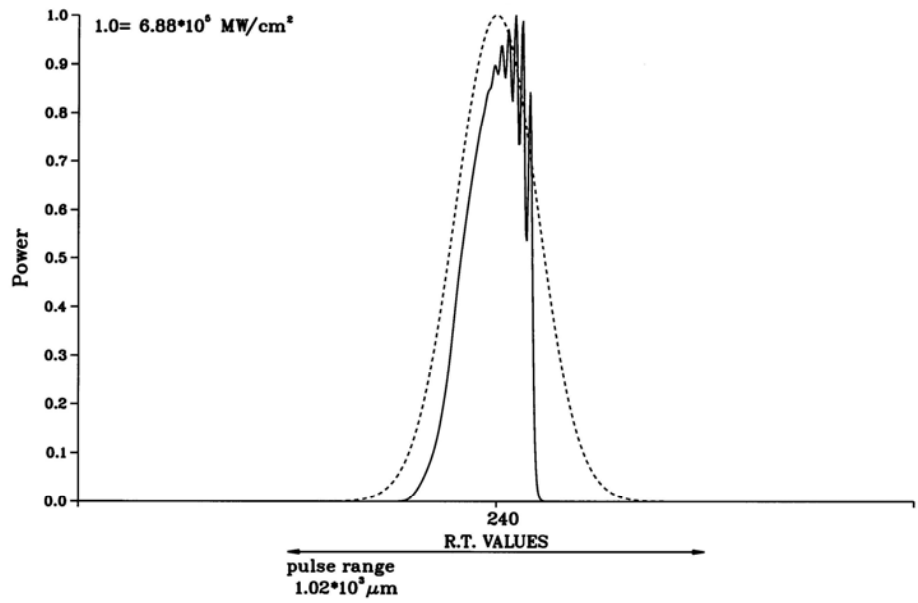

**Fig. 16c**

**Fig. 16** Comb-like structure in FEL oscillators in deep saturation (round trip 240) at zero cavity mismatch $g_0 = 1, \quad \eta = 0.06$; a) $\pi \mu_T = 0$, b) $\pi \mu_T = -15$, c) $\pi \mu_T = 15$ (dotted curve represent the electron longitudinal distribution)

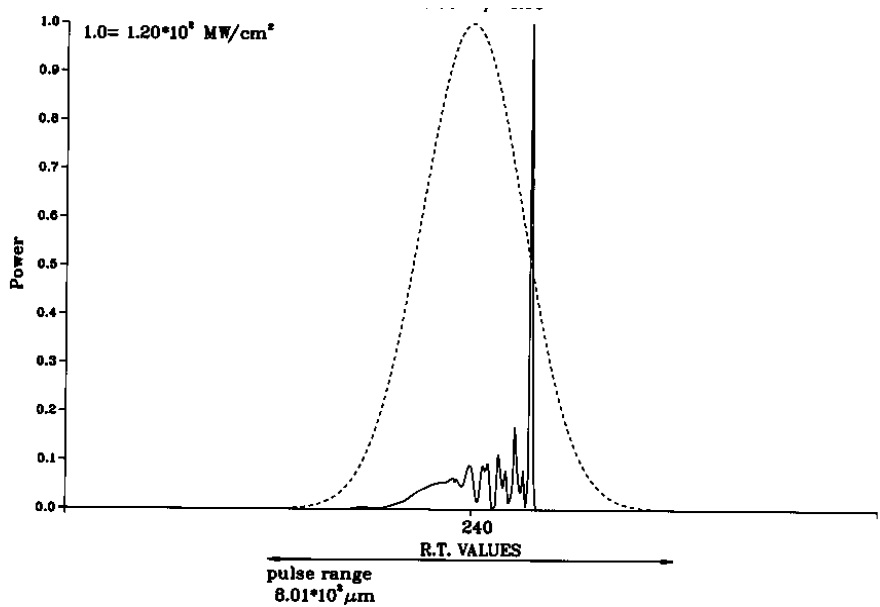

**Fig. 17a**

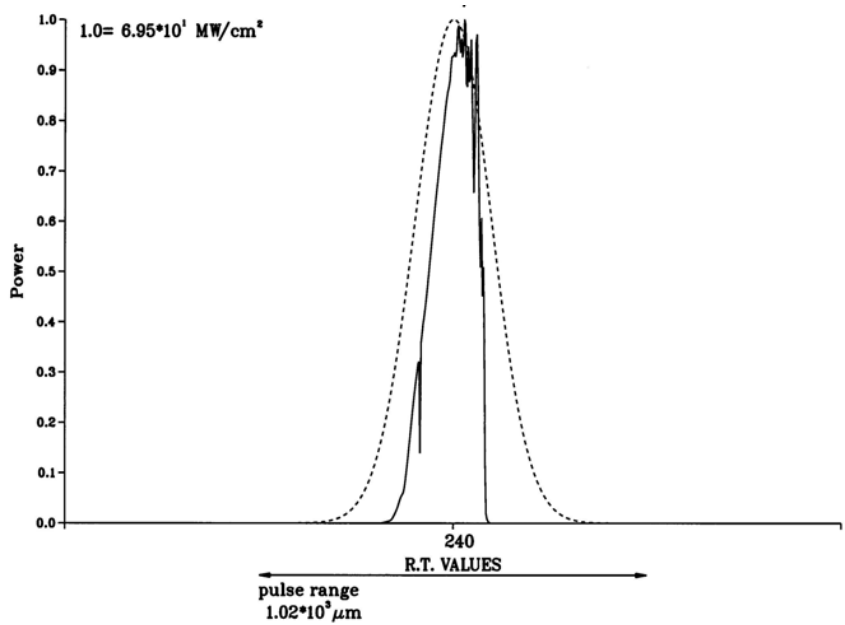

**Fig. 17b**
**Fig. 17** Third Harmonic in deep saturation, same parameters as Fig. 16,
a) $\pi \mu_T = 0$, b) $\pi \mu_T = 15$

A further key point is the role of the lethargy, which we have just touched on in the paper. Plots reported in Fig. 18 can be considered clarifying: the figure shows the behavior of the stable intracavity power vs. the cavity length mismatch $\delta = \frac{\Delta L}{\lambda}$, where $\Delta L$ is the cavity length reduction necessary to compensate the lethargy effect and $\lambda$ the operating wavelength. It is evident that the condition of reverse tapering yields the best conditions of operations, in terms of power but not for the stability curve. The region of cavity mismatch allowing the growth of stable output power is larger in the case of zero tapering[4]. This is just a consequence of the fact that the gain curve is narrower in the case of reverse tapering. The case of $\pi \mu_T = -15$ is evidently the less favorable.

---

[4] In a forthcoming investigation we will quantify more precisely these effects,

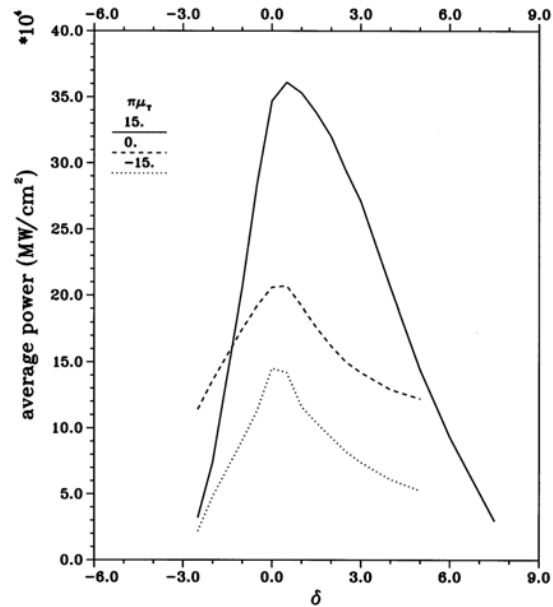

**Fig. 18** Equilibrium intracavity power intensity vs. $\delta$, $g_0 = 1$, $\eta = 0.06$.
The average power is the average on laser packet distribution

Regarding the harmonics we have reported in Fig. 19 the round trip power evolution of the first, third and fifth harmonic for $\pi\mu_T = 0, \pm 15$. The effect of the tapering on the higher harmonics is quite remarkable, if confronted to the case with no tapering. In this last case the harmonic power exhibits a bump before that the fundamental has reached the onset of the saturation and then decreases even by an order of magnitude. When tapering is active the power of the higher order harmonics remain almost constant, even when the fundamental has deeply saturated. The case with inverse tapering yields a more efficient result in terms of harmonic power too.

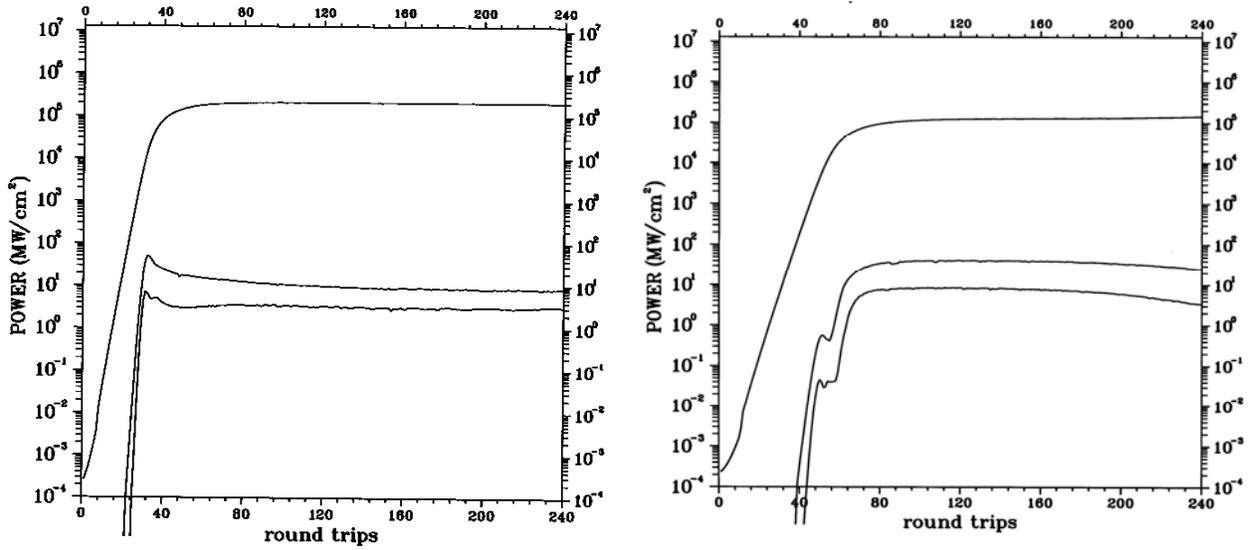

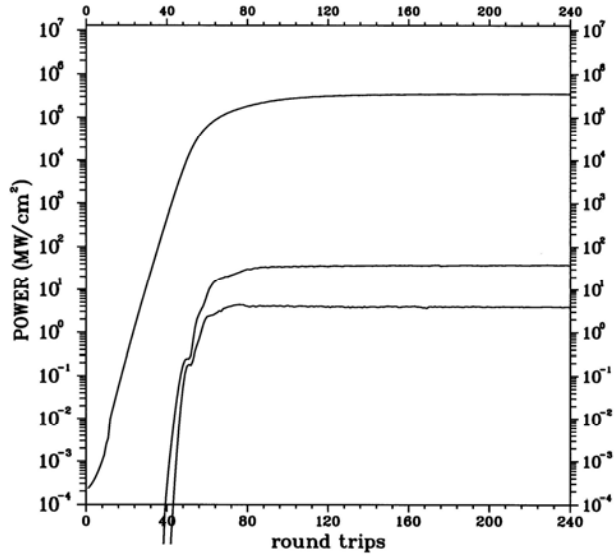

**Fig. 19** Power evolution of the fundamental, third and fifth harmonic at zero cavity mismatch; a) $\pi\mu_T = 0$; b) $\pi\mu_T = -15$; c) $\pi\mu_T = 15$

The final point we will treat is relevant to the case of the pulse dynamics for values of the tapering parameter giving rise to the gain instability reported in the previous sections. We have already remarked that in this region there are two competing modes which may give rise to a quite interesting dynamics and indeed we may have two different carrying frequencies separated by $\delta = \dfrac{\omega_2 - \omega_1}{\omega_{res}} \cong \dfrac{1}{N}$.

The full dynamics is rather difficult to check because the code becomes time consuming since a large number of competing modes should be included in the calculation and perhaps the slowly varying amplitude approximation is not valid any more. We have however considered the mode dynamics of a single carrying frequency with slightly larger gain with respect to the other. For such large values of the tapering, the system reaches the saturation after a large number of round trips, as it should be, since the gain is quite low, even for large values of the small signal gain parameter (see Fig. 20)

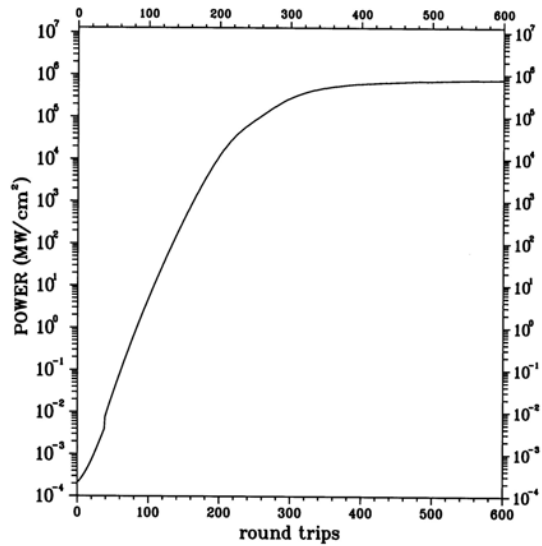

**Fig. 20** Intracavity power growth at zero cavity mismatch, $g_0 = 1$, $\pi\,\mu_T = 25, \eta = 4\%$

We have noted that, at zero mismatch, the slippage effect is no more active and that the optical field remains locked to the electron bunch in an almost congealed position over the entire evolution which in our simulation went well beyond the onset of the saturation (see Fig. 21). Beyond the saturation point the optical pulse becomes distorted and small structures over the peak of the bunch start to develop.

Results of this paper can be summarized as follows:

a) FEL oscillators, operating with undulators having a uniform tapering, exhibit an interesting behavior associated with the peculiar nature of the gain function.

b) The uniform tapering guarantees an enhancement of the efficiency which is not the result of an optimization criterion as it happens in the case of the amplifier.

c) The pulse propagation dynamics displays a very interesting phenomenology, which indicates that an interesting interplay may occur between tapering, slippage and lethargy. We have got some indication that the combined use of these effects can be useful to model the pulse shape.

In a forthcoming investigation we will explore more deeply the pulse propagation dynamics in tapered oscillators and show that definite advantages (in terms of efficiency, pulse shapes…) can be obtained with undulators with non uniform tapering.

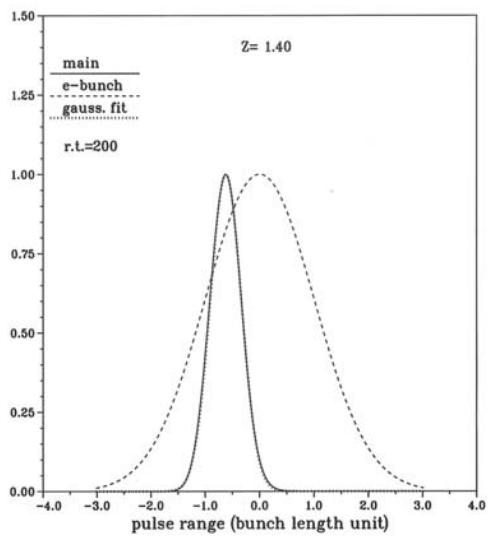 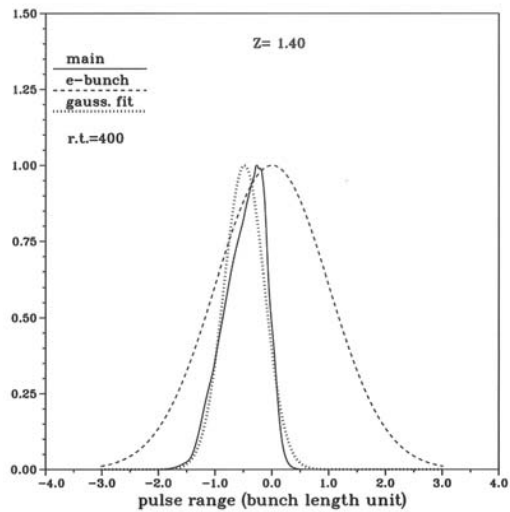

**Fig. 21a**　　　　　　　　　　**Fig. 21b**

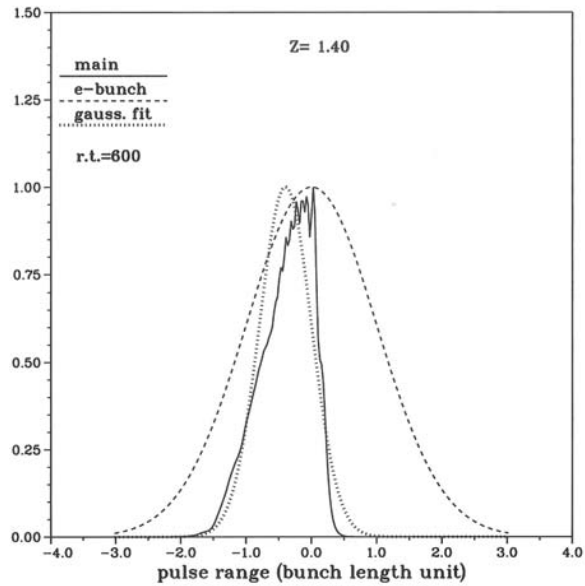

**Fig. 21c**

**Fig. 21** Optical pulse shapes at different round trips for $g_0 = 1,\ \pi\mu_T = 25, \eta = 4\%$

a) round trip 200, b) round trip 400, c) round trip 600.

### ACKNOWLEDGMENTS

The Authors express their sincere recognition to Dr. S. Benson for clarifying discussions on the tapered FEL and to Dr. Luca Giannessi for confirming the existence of the pulse comb-like structure in deeply saturated FEL oscillators with the code Perseo.

"Introduction to the Physics of Free Electron Laser and Comparison with Conventional Laser Sources" http://arxiv.org/abs/1010.1647v1 (submitted for publication)